# On the Safety Loading for Chain Ladder Estimates: A Monte Carlo Simulation Study

M. Schiegl

**Abstract**

A method for analysing the risk of taking a too low reserve level by use of Chain Ladder method is developed. We give an answer to the question of how much safety loading in terms of the Chain Ladder standard error has to be added to the Chain Ladder reserve in order to reach a specified security level in loss reserving. This is an important question in the framework of integrated risk management of an insurance company. Furthermore we investigate the relative bias of Chain Ladder estimators. We use Monte Carlo simulation technique as well as the collective model of risk theory in each cell of run-off table. We analyse deviation between Chain Ladder reserves and Monte Carlo simulated reserves statistically. Our results document dependency on claim number and claim size distribution types and parameters.

1. Introduction

Claims reserving is a very important topic for P&C insurance companies for all lines of business with long run-off period. Therefore a variety of mathematical methods for estimation of reserves (total loss amounts) has been developed (Institute of Actuaries 1997). One of the well known and frequently used methods is the Chain Ladder (CL) method (Mack 1997). An analytic expression for the standard error of the CL reserve was proposed by Mack (1993).

The present paper has two aims: First it looks for relative bias in CL reserves. We call the difference between CL reserve and Monte Carlo reserve ("real reserve") divided by the square root of m.s.e. (Section 2) "relative bias". The second aim is to detect a relation between the CL reserve's standard error and the risk of taking a too low reserve level. This should give an answer to the question of how much safety loading in terms of the CL standard error (as proposed by Mack (1993)) has to be added to the CL reserve in order to reach a specified security level in loss reserving.

Further we establish a method which allows to give an answer to those questions for every given loss model. In the present paper we concentrate on a special stochastic model: For each cell in run-off triangle we apply the collective model of risk theory with Poisson, binomial or negative binomial claim number process as well as Pareto or exponential single claim size process. As a limiting case of the collective model we investigate the pure claim number process (size of each claim is one currency unit), as for this case equivalence between CL estimator and maximum likelihood estimator was proven by Schmidt and Wünsche (Schmidt and Wünsche 1998) for all three types of claim number distribution referred to in this paper. The result on maximum likelihood estimator was published earlier in the Poisson case (Hachemeister and Stanard 1975; Mack 1991, 1997). We find our results by a computational study via Monte Carlo (MC) simulation techniques.

The paper is organised as follows: In Section 2 we give a short introduction to claim reserving of P&C companies and introduce the CL method as a mathematical reserving technique. The stochastic model used in the present paper is defined in Section 3. Section 4 describes the backtesting concept which is the basis for answering the above mentioned questions. Its application is documented by introducing the structure of MC simulation code. Furthermore tests of the code and its test results are discussed. Results of our investigations are given in Sections 5 and 6. Section 5 deals with results of pure claim number model while Section 6 focuses on total claim size model. Finally Section 7 gives the conclusions of this paper.

## 2. Run-off Triangle and CL Method

In practice two kinds of Reserves have to be estimated: IBNR Reserves for already incurred damages which have not jet been reported and IBNER for cases which have been reported but not enough reserved. In order to apply estimation techniques one has to classify historical data of pay outs and / or total claim sizes for a statistically significant part of an insurance portfolio according to occurrence year, reporting year or development year. The type of the representation used depends on the aim of the calculation and on the available data. An overview of loss reserving is given in (Mack 1997, Taylor 1986, 2001).

A two dimensional representation of data as a basis for mathematical estimation methods is called run-off triangle: Let $S_{ik}$ be the (incremental) claim amount paid in development year k, and hence in calendar year i + k, for claims occurred in year i.

|  | **Development Year →** | | | | | | |
|---|---|---|---|---|---|---|---|
|  | 1 | 2 | 3 | 4 | 5 | 6 | 7 |
| 1 | $S_{11}$ | $S_{12}$ | ... | $S_{1k}$ | ... | $S_{1, I-1}$ | $S_{1I}$ |
| 2 | $S_{21}$ | $S_{22}$ | ... | $S_{2k}$ | ... | $S_{2, I-1}$ |  |
| 3 | ... | ... |  | ... |  |  |  |
| 4 | $S_{i1}$ | $S_{i2}$ | ... | $S_{ik}$ |  |  |  |
| 5 | ... | ... |  |  |  |  |  |
| 6 | $S_{I-1, 1}$ | $S_{I-1, 2}$ |  |  |  |  |  |
| 7 | $S_{I1}$ |  |  |  |  |  |  |

↓ Occurrence Year

Calendar years stand along the diagonals. The upper part of the table is filled with data concerning the past. For each new business year the table grows along its diagonal. The lower (empty) part of the table represents the future. The values have to be estimated with the help of an appropriate method. The sum of all those future claim amounts is the total reserve. The reserve for occurrence year i can be calculated as:

$$R_i = S_{i,I+2-i} + S_{i,I+3-i} + ... + S_{iI}, \quad (i = 2, ..., I)$$

which is just the sum of the estimated columns per row.

One of the well known methods for reserve estimation is the CL method (Mack 1997). In the following section a short review of the method is given in summarizing the formulas used in our calculations.

Cumulative claims are defined as

$$C_{ik} = \sum_{l=1}^{k} S_{il} \ .$$

CL factors are defined as

$$\hat{f}_k = \left. \sum_{i=1}^{I-k} C_{i,k+1} \middle/ \sum_{i=1}^{I-k} C_{ik} \right. \ .$$

Estimators for the lower part of the run-off triangle are given by

$$\hat{C}_{ik} = C_{i,I+1-i} \cdot \prod_{l=I+1-i}^{k-1} \hat{f}_l \ .$$

The estimator for the reserves in the cumulative representation can be written as

$$\hat{R}_i = \hat{C}_{iI} - C_{i,I+1-i} \ .$$

The estimator for the total Reserve is

$$\hat{R} = \sum_{i=2}^{I} \hat{R}_i \ .$$

The mean square error (m.s.e.) is a measure for the future volatility of expected claim amounts and is defined for occurrence year i as

$$m.s.e(\hat{R}_i) \equiv E((R_i - \hat{R}_i)^2 / D)$$

where D is a symbol for the (fixed) run-off-triangle.

In ( Mack 1993 ) the following estimator for the m.s.e. is proposed:

$$m.s.e(\hat{R}_i) = \hat{C}_{iI}^2 \sum_{k=I+1-i}^{I-1} \frac{\hat{\sigma}_k^2}{\hat{f}_k^2} \left( \frac{1}{\hat{C}_{ik}} + \frac{1}{\sum_{j=1}^{I-k} C_{jk}} \right)$$

where

$$\hat{\sigma}_k^2 = \frac{1}{I-k-1} \sum_{j=1}^{I-k} C_{jk} \left( \frac{C_{j,k+1}}{C_{jk}} - \hat{f}_k \right)^2$$

and the mean square error for the total Reserve is given as (Mack 1993):

$$m.s.e(\hat{R}) = \sum_{i=2}^{I} m.s.e(\hat{R}_i) + \sum_{i=2}^{I-1} \left[ \hat{C}_{iI} \left( \sum_{j=i+1}^{I} \hat{C}_{jI} \right) \sum_{k=I+1-i}^{I-1} \frac{2\hat{\sigma}_k^2 / \hat{f}_k^2}{\sum_{n=1}^{I-k} C_{nk}} \right]$$

The answer to the following question which is of great practical importance is however open: What is the risk of reserving a too small amount (underreserving risk) by setting the reserve to a fixed value? This is a typical question that arises in any insurance company. We try to find an answer to that question by analysing empirical probability distributions of the stochastic variable $\delta = (R_{CL} - R_{real})/\sqrt{m.s.e}$. Where $R_{CL}$ and m.s.e. are the expected value and mean square error (the square of standard error) of total reserve according to CL method. $R_{real}$ is the MC simulated reserve, a realisation of the real reserve within the framework of used stochastic model (see Sections 3 and 4). In this way we gain information about the underreserving risk, i.e. the deviation between the reserve calculated by CL method and the real reserve in terms of standard error. We call $E[\delta]$ "relative bias". Notice that the bias $E[R_{CL} - R_{real}]$ and the relative bias can have different signs, as

$$E[\delta] = E[R_{CL} - R_{real}] E\left[ \frac{1}{\sqrt{m.s.e.}} \right] + Cov\left[ R_{CL} - R_{real}, \frac{1}{\sqrt{m.s.e.}} \right].$$ For simulating run-off tables an appropriate model for claim generation is needed. In the following we introduce the model used in this paper.

## 3. The Model

The model assumes that all incremental claims $N_{ik}$ are independent. The ultimate aggregate claim number $N_i = \sum_{k=1}^{I} N_{ik}$ of occurrence year i is Poisson, negative binomial or binomial distributed with equal distribution type for each year.

A Poisson distributed variable $\sim \Pi(\lambda)$ has the density

$$p_n = e^{-\lambda} \frac{\lambda^n}{n!},$$

a binomial distributed variable $\sim B(m, p)$ has the density

$$p_n = \binom{m}{n} p^n (1-p)^{m-n},$$

a negative binomial distributed variable $\sim NB(\rho, p)$ has the density

$$p_n = \binom{\rho + n - 1}{n} p^\rho (1-p)^n$$

The incremental claim numbers $N_{ik}$ are defined as the number of claims contributing to the claim amount $S_{ik}$ (see Section 2). The ultimate aggregate claim number $N_i$ for occurrence year i is distributed over the I development years according to the multinomial distribution $MN(N_i, \vec{\pi})$ with a given run-off pattern $\vec{\pi}$ with elements $\pi_k$, $k \in \{1, 2, \ldots, I\}$. The distribution of incremental claim numbers $N_{ik}$ is given by

$$f(N_{i1}, N_{i2}, \ldots, N_{ik}, \ldots, N_{iI}; \vec{\pi}) = \frac{N_i!}{\prod_{j=1}^{I} N_{ij}!} \prod_{j=1}^{I} \pi_j^{N_{ij}}$$

with $\sum_{j=1}^{I} \pi_j = 1$ and $\sum_{k=1}^{I} N_{ik} = N_i$.

In the case of claim number simulation the $S_{ik}$ (as defined in Section 2) are set equal to $N_{ik}$. This means that the single claim amount is assumed to be one currency unit and therefore CL method is applied to claim numbers only.

In the case of total claim size model we simulate claim amounts $S_{ik}$ in each cell of the run-off table according to the collective model. Therefore the incremental claim amount is given by

$$S_{ik} = \sum_{l=1}^{n_{ik}} X_l$$

where $X_l$ is the single claim amount and $N_{ik}$ the number of claims in development year k occurred in year i. $N_{ik}$ is a random variable as defined above. We assume $X_l$ to be Pareto distributed with density

$$f_{r,\alpha}(x) = \begin{cases} \dfrac{\alpha-1}{r^{1-\alpha}} x^{-\alpha} & \text{for } x > r \\ 0 & \text{for } x < r \end{cases}$$

with $E[X_l] = \dfrac{\alpha-1}{\alpha-2} r$ and $E[X_l^2] = \dfrac{\alpha-1}{\alpha-3} r^2$.

or exponentially distributed with density

$$f_\mu(x) = \begin{cases} \mu \cdot e^{-\mu(x-r)} & \text{for } x > r \\ 0 & \text{for } x < r \end{cases}$$

with $E[X_l] = r + \dfrac{1}{\mu}$ and $E[X_l^2] = \left(r^2 + \dfrac{2r}{\mu} + \dfrac{2}{\mu^2}\right)$.

Parameter r represents the minimum claim size (cut off size for heavy losses). The Pareto distribution loses its first moment for $\alpha <= 2$ and its second moment for $\alpha <= 3$. In this paper we investigate the interval $2.1 <= \alpha <= 4$.

<u>Comments on the Model:</u>

In the case of Mack's model the following is assumed:

(1) The $C_{ik}$ of different occurrence years are independent.

(2) $E[C_{ik+1} | C_{i1},...,C_{ik}] = C_{ik} f_k$

(see chapter 3.2.4. of (Mack 1997))

In this case the CL estimators are known to be unbiased.

The model we use in the present paper contradicts of course these assumptions. On the other hand numerical investigations are justified due to this.

The model we use in the present paper belongs to the cross classified parametric models which are known from literature (see chapter 3.3. of (Mack 1997)). The collective model as a generating process for the incremental claims $S_{ik}$ is treated there, too. For these models a weaker assumption than the CL assumption holds: $E[C_{ik+1}] = E[C_{ik}]f_k$. This is obviously a generalization of assumption (2) of Mack's model.

The reason why we apply the CL method to this class of models are twofold: First, literature is concerned with this kind of problems (Mack 1997; Schmidt and Wünsche 1998). Second; in practice the CL method is used also in cases where one can not be sure that the assumptions of Mack's model hold. We intend to introduce a numerical backtesting method for such cases.

In Mack's model the use of the proposed m.s.e. estimator is connected to a third assumption:

(3) $Var[C_{ik+1}|C_{i1},...,C_{ik}] = C_{ik}\sigma_k^2$

(see chapter 3.2.5. of (Mack 1997))

We use the proposed estimator for the m.s.e. as a normalization factor for differences of reserves in order to discuss about relative quantities and not absolute ones.

4. The Monte Carlo Simulation Method – Description of Algorithm and Tests

In Section 2 we defined a stochastic variable $\delta := (R_{CL} - R_{real})/\sqrt{m.s.e}$ which has to be simulated via MC techniques. We describe the structure of the used MC code in the following:

We MC simulate $n_{r.o.}$ different run-off scenarios (usually $n_{r.o} = 1000$) according to the following steps:

1) Generate the ultimate aggregate claim number for each of the I occurrence years.
2) Distribute the ultimate aggregate claims according to the multinomial distribution to receive the $N_{ik}$ claims in cell (i, k).

3) In the case of total claim size simulation: Generate for each cell of the run-off table $N_{ik}$ Pareto or exponentially distributed random variables and compute $S_{ik}$. In the case of claim number simulation set $S_{ik}= N_{ik}$.

4) Calculate the cumulated run-off triangle, the $C_{ik}$ as defined in Section 2 as basis for the CL method.

5) Calculate the CL reserve $R_{CL}$ and the mean square error m.s.e according to CL method (Section 2) from the upper part (triangle) of the simulated run-off table.

6) Calculate $R_{real}$ from simulation of run-off table's lower part

7) Calculate $\delta = (R_{CL} - R_{real})/\sqrt{m.s.e}$

We repeat steps 1) – 7) for each of the $n_{r.o.}$ different run-off scenarios. We calculate the empirical distribution function of the variable $\delta$ from these realizations of the variable as well as the expected value of $\delta$, $E(\delta)$. To estimate the statistical errors we repeat the procedure 10 times and calculate the mean and standard error in each point of the empirical distribution function and $E(\delta)$. This gives us a crude estimate of statistical stability of the measured quantities.

To make sure that the simulation delivers results which are correct and obey the model which is introduced in Section 2 we perform several tests: We test the correctness of the several used random generators (Poisson, binomial, negative binomial, Pareto, exponential) and the multinomial distribution of claims along development years. Additionally we found convergence of results as a function of the number of MC samples, as it has to be.

We want to demonstrate the results of one test as an example representing all the others: Claim amount S in each cell of the run-off table is MC simulated according to the collective model: $S = \sum_{i=1}^{N} X_i$. We compare the empirical (MC) S – distribution function with the result of a Panjer recursion in the Poisson case: According to Panjer the discretised distribution of S is given by:

$$g_k \equiv P(S = k \cdot h) = \sum_{j=1}^{k} \lambda \frac{j}{k} f_j g_{k-j}$$

where $\lambda$ is the Poisson parameter, h the width of a (discretisation) step and $f_i$ the discretised density of the random variable X. Figure 1 a) and b) shows the results of 10 different MC simulation (each 1000 realizations of S), mean (black diamonds) and mean +/- standard deviations (black crosses) as well as the Panjer recursion distribution function (black solid line). The single claim sizes $X_i$ are Pareto distributed. Figures 1 a (b) have the following parameters: $\alpha$ = 4 (2.1) / $\lambda$ = 12 / r = 1000 / h = 5 (50). Obviously there is an excellent agreement between MC and Panjer results.

5. Results for the Claim Number Model

In the following we present results of MC simulation for claim number model as described in Section 3. The model we use has been treated in literature (Schmidt and Wünsche 1998; Hachemeister and Stanard 1975; Mack 1991, 1997). Equality between CL estimators and maximum likelihood estimators for that special model was shown there.

We perform simulations with different sets of parameters: We vary length of run-off period ( l = 5, 10, 15, 20 years ), run-off pattern (linear or exponential; values in use for $\pi_t$ see table 1), the type of ultimate aggregate claims' distribution function (Poisson, binomial, negative binomial) and distribution parameters of these in order to investigate different expected claim numbers and different standard deviations of claim numbers. We found the following results:

(i)        In all MC simulations the 50% percentile of the $\delta$ - distribution lies at $\delta = 0$ whereas the expected value of $\delta$ is in general smaller than 0. This means that CL method produces a reserve which in 50% of the cases is too high and in 50% too low. While for positive signs of $\delta$ ($R_{CL} > R_{real}$, positive bias), $\delta$ has a high affinity to smaller values ("high m.s.e."), for negative signs of $\delta$ ($R_{CL} < R_{real}$, negative bias), $\delta$ has high affinity to more negative values ("low m.s.e."). This is of course no contradiction to literature (Schmidt and Wünsche 1998) because maximum likelihood estimators are not necessarily unbiased.

(ii) The width of $\delta$ - distribution and the deviation of the expected value from zero depend on the length of the run-off period I. In figure 2 we see percentiles of $\delta$ - distribution function as well as the expected value of $\delta$ ( $E(\delta)$ ) as a function of the run-off period I. We use Poisson distributed ultimate aggregate claim number with $\lambda = 100$ (per occurrence year) and exponential as well as linear run-off pattern. Depicted are the 5%, 10%, 50%, 90% and 95% percentiles. In all four parts of the figure we show the mean of the above mentioned 10 (equally parametrised) MC simulations and the regions of plus or minus one standard deviation of the 10 results as black bars. With increasing run-off period both the width of distribution as well as the $E(\delta)$'s deviation from zero decreases. We can read from figure 2: For I = 5 the difference in expected value between the real reserve and the CL reserve is more than 8% of (CL) standard error for the linear pattern (more than 5% for the exponential pattern) and decreases to some 3% of the standard error for I = 20. We further observe that changes (due to run-off period) in $\delta$-distribution as well as in $E(\delta)$ decreases for increasing I. We compare the size of changes with the width of error bars to decide between effects and "statistical noise". We therefore find that quality of CL estimates increases with increasing length of run-off period and seems to come to a point of saturation at about 15 years. We observe only slight differences between linear and exponential run-off patterns: For linear pattern and I = 5 a broader $\delta$ - distribution with higher $E(\delta)$ deviation from zero compared with the results of exponential pattern is detected (see figure 2). This situation improves more rapidly in case of linear pattern as I increases than it does for the exponential case: For I = 15, 10 and 20 the results for both patterns are comparable (see also (v) ). We find similar results for Poisson claim number distribution with $\lambda = 50$.

(iii) In the Poisson case we find a slight dependence on the expected claim number $\lambda$. For $\lambda = 100$ $\delta$ - distribution is a bit broader, but less asymmetric and therefore E($\delta$) is closer to zero compared with $\lambda = 50$. More obvious is the fact, that for $\lambda = 100$ E($\delta$) converges better to zero for increasing run-off period I than it does for $\lambda = 50$. (Compare figure 2 with table 2a.) Variation of I (see (ii)) has much greater impact on results than variation of $\lambda$.

(iv) Having in mind the last two topics, asymptotic convergence of E($\delta$) to zero seems possible for increasing run-off period and expected claim numbers. Expected value of claim number ($\lambda$ in Poisson case) is however not the appropriate parameter for convergence considerations, because this is the expected ultimate aggregate claim number for one occurrence year. I different occurrence years are accumulated in a run-off table. This influences convergence properties. We show that I and $\pi$ have strong impact on convergence ( see (ii) and (v) ). Therefore an appropriate combination of E(N), I and $\pi$ has to be found to discuss convergence topics. We investigate convergence as a function of $\Delta := E(N) \cdot \sum_{j=1}^{I} (I + 1 - j)\pi_j$ in the Poisson case:

$\Delta := \lambda \cdot \sum_{j=1}^{I} (I + 1 - j)\pi_j$ . $\Delta$ is the expected value of claims in the run-off triangle (upper / known part of table). We have chosen this definition for $\Delta$, because it depends on the above mentioned important parameters and has additionally an practical interpretation: The number of claims the run-off table is "made" of. For comparison: $E(N) \cdot I$ is the expected value of claims in the total run-off table.

Figure 3 shows E($\delta$) as a function of $\Delta$. Depicted are E($\delta$), the mean of 10 MC simulations (black diamonds) and the region of one standard deviation below and above the mean (sketched line). All results of our Poisson case simulations are included in the figure; this means different values of I, $\pi$ and $\lambda$. In this sense figure 3 shows a global result: E(δ) converges asymptotically to 0 for increasing Δ. The end of convergence is reached at $\Delta \approx 5000$.

(v) We observe a small, but within the given range of error barrs noticeable dependence of δ - distribution on the type of run-off pattern in the case of short run-off period (see figure 2). The width of δ - distribution is a measure for quality of CL estimates in the framework of that special model. The wider the distribution the less secure are CL predictions. We find distributions for linear run-off patterns and I = 5 a bit wider than for exponential ones. This is especially true for the left tail of the distribution. For longer run-off periods (10, 15, 20 years) there is not much difference between the exponential and linear pattern. We interpret this result as follows: CL method is slightly superior in estimating exponential run-off patterns compared to linear ones for short run-off periods. This discrepancy vanishes for increasing run-off periods.

(vi) The δ - distribution as well as E(δ) do not significantly depend on the type of claim number distribution (Poisson, binomial, negative binomial) provided that distribution parameters have been chosen to fit first moment. We find also that there is not much difference in δ - distribution and E(δ) if we chose different standard deviations of claim number distribution at fixed expected value. For a selection of a few examples see figure 4. There are depicted percentile representations of δ - distribution for different types of claim number distributions; all with expected value (claim number) very close to 50 and a wide range of different variances for I = 5 and linear run-off pattern. We find similar behaviour for I = 15, exponential run-off pattern and other expected values (without figure).

6. Results for the Aggregate Claims Model

As described in Section 3 we do not only deal with claim number model but also with total (aggregate) claim size model according to the collective model of risk theory as an enhancement of the first. The claim number model can be interpreted as a special case of total claim size model where each claim is cut at one unit of currency. Now we turn to the results of total claim size model's MC simulations. In analogy to Section 5 we vary length of run-off period (I = 5, 10, 15, 20 years), run-off pattern (linear and exponential, see table 1) and parameters of single claim size distribution. We focus here on Poisson distributed (ultimate aggregate) claim numbers and vary the parameter $\lambda$.

(i) As in Section 5 (i) we observe here: The 50% percentile of $\delta$ - distribution function is very close to 0 and $E(\delta) < 0$ in all MC simulations performed.

(ii) We discuss the dependence of $\delta$ - distribution on the Pareto exponent $\alpha$. Figure 5 shows the 5%, 10%, 20%, 50%, 80%, 90% and 95% percentile of $\delta$ - distribution as a function of $\alpha$ in a region from $\alpha = 2.1$ to $\alpha = 4$. The parameters are chosen as follows: I = 20 (a) and I = 5 (b); exponential run-off pattern; $\lambda = 100$; r = 1000. One can see that for $\alpha > 3.5$ $\delta$ - distribution is independent of $\alpha$. For $3.0 < \alpha < 3.5$ we observe a slight dependency. For $\alpha < 3$ where the single claim size distribution loses its second moment (transition to "uninsurable" risk), there is a very clear shift of percentiles smaller than 50% to more negative values. The 50% percentile is close to 0 for all $\alpha$. Obviously there is a left skew (asymmetric $\delta$ - distribution density) which is accentuated drastically for $\alpha < 3$. We performed several other MC simulations to investigate the $\alpha$-dependence. We find comparable behaviour independent of the type of run-off pattern as well as the claim number process parameters. From these observations we draw several conclusions: CL reserve is too low (too high) with an (equal) probability of 50%. But we find $E(\delta) < 0$ for all simulations. This effect is drastically enhanced for small $\alpha$. Therefore the risk of taking a too small reserve level is increased with decreasing α (especially for single claim size distribution with infinite second moment). With a given security level calculations as performed here can help the actuary to set reserves: For example an insurance company wants to carry a risk of only 5% of having insufficient reserves and the Pareto exponent of single claim size distribution has been measured to be 2.1 for a portfolio with I = 20. Then one can read from figure 5 a), that the reserve has to be set to a value equal to CL reserve plus 4.5 times the square root of m.s.e. as defined in Section 2. (Remember: δ has been defined as $\delta := (R_{CL} - R_{real})/\sqrt{m.s.e.}$ ; see Section 4). If the company accepts a underreserving risk of 10%, 2.6 times the standard error in addition to CL reserve is sufficient. For 5% risk but Pareto exponent α = 3.5 reserve has to be set to the value: $R_{CL} + 1.9 \cdot \sqrt{m.s.e.}$ . Also the following conclusions can be drawn from figure 5: The smaller the risk level for underreserving, the higher the safety loading that has to be added to the reserve and: The smaller α, the higher the safety loading has to be chosen for fixed risk level. Both observations are intuitively clear and agree with common reserving practice. The fact that underreserving risk (measured in units of $\sqrt{m.s.e.}$ ) is drastically increased for α < 3 is consistent with the definition of m.s.e (see Section 2) which is based on a second moment concept. It is therefor appropriate to measure safety loading in units of $\sqrt{m.s.e.}$ only for distributions with finite second moment (see figure 5: for α > 3.5 safety loading nearly independent on α). For distributions without second moments other concepts for safety loading have to be used.

(iii) For all simulations presented in this paper the second Pareto parameter is kept at r = 1000. Variation of r does not influence results, as we investigate δ 's distribution and expectation which are relative quantities.

(iv) Concerning the dependence on run-off period I we find for total claim size model a corresponding behaviour to pure claim number model: The width of δ - distribution is accentuated for decreasing run-off period I as shows a comparison between figure 5

a) (I = 20 years) and b) (I = 5 years). The $\delta$ - distribution function at $\alpha > 3.5$ is determined by claim number model as shows a comparison between figure 5 a) b) and the upper left part of figure 2. (Compare Fig. 5a (b) $\alpha = 3.5$ or 4.0 to Fig. 2 upper left part for I = 20 (I = 5).) For smaller $\alpha$ ("uninsurable" risks with infinite second moment in single claim size distribution) the $\delta$ - distribution function is strongly influenced by single claim size distribution.

(v) Figure 6 shows E($\delta$) as a function of $\Delta$ (as defined in Section 5) for the total claim size model (compare figure 3 and Section 5 (iv) for claim number model). Figure 6 a (b) shows the results for single claim size distribution parameter $\alpha = 4.0$ ($\alpha = 2.1$). Depicted are E($\delta$), the mean of 10 MC simulations (black diamonds) and the region of one standard deviation below and above the mean (sketched line). In this figure different parameter sets for Poisson distributed claim numbers are combined. This means different values of I, $\pi$ and $\lambda$. In this sense the depicted results are global. For $\alpha = 4.0$ absolute value of E($\delta$) decreases with increasing $\Delta$ but in contrast to pure claim number model (see figure 3) E($\delta$) does not approach zero at $\Delta$=5000 (figure 6 a). There is a reduction of rate of convergence at $\Delta \approx 1500$. But at $\Delta = 5000$ we find still a deviation of 4 – 5% of CL standard error from zero. Additionally to the pure claim number process we model here a claim size process which brings about higher volatility of $\delta$ and disables convergence of E($\delta$). It can not be decided with current results, if there is definitely no asymptotic convergence to zero or if there is one at $\Delta \gg 5000$ (see figure 6 a). As can be expected from E($\delta$)'s behaviour, for percentiles of $\delta$-distribution function we observe a saturation effect at $\Delta \approx 1500$. Percentile values for $\Delta \approx 5000$ can be found in table 2b. This means that the higher the number of claims in run-off table, the less safety loading (in units of CL standard error) that needs to be added to a CL reserve to reach a given level of underreserving risk. But this risk reduction mechanism becomes inefficient for $\Delta > 1500$ because further increase in $\Delta$ doesn't bring about further reduction of safety loading. For single claim size parameter $\alpha$ = 2.1 we don't observe any sign of asymptotic convergence of E($\delta$) as is obvious from figure 6 b). E($\delta$)'s deviation from zero lies around 100% - 150% of CL standard error independent of the size of $\Delta$. $|E(\delta)|$ is therefore more than one order of magnitude higher than in the former case ($\alpha$=4.0). This is plausible due to the fact that single claim size distribution has ceased to have a finite second moment (for $\alpha$=2.1) and we are dealing with "uninsurable" risks with infinite standard deviation. This leads obviously to a strongly asymmetric $\delta$-distribution density with high down side risk for insufficient reserves. As the $\delta$-distribution function is concerned, we find similar behaviour as before: A saturation effect at $\Delta \approx 1500$, of course at a higher level. For values see table 2b.

(vi) We investigate the influence of single claim size distribution function on E($\delta$) and $\delta$-distribution function. We compare Pareto with exponential distributed claim sizes (parametrisation as defined in Section 3) for two different sets of parameters. We have to adjust the two different types of single claim size distribution functions. The <u>two</u> free parameters of exponential distribution have to be chosen in an appropriate way. To generate equivalent measures we use the same value for the cut off claim size r for both types of distribution. We chose the second parameter, $\mu$, in a way that the first moments of both types of distributions coincide. As there are only <u>two</u> free parameters no higher moments can be taken into consideration. This kind of fit procedure is of course somehow arbitrary. Therefore we tested an other one: A least square fit procedure on the distribution densities with fixed r and variable $\mu$. This change had no substantial influence on our results. This is the reason why we stay at our first choice. We intend to find out differences in results with the following 2 couples of differently parametrised claim size distributions: Pareto ($\alpha$=4.0 / r=1000 / $m_1$=1500 / STDV = 866), exponential ($\mu$=0.002 / r=1000 / $m_1$=1500 / STDV = 500) on the one hand and Pareto ($\alpha$=3.1 / r=1000 / $m_1$=1909.09 / STDV = 4166), exponential ($\mu$=0.0011 / r=1000 / $m_1$=1909.09 / STDV = 909.09) on the other. To the first couple

we will refer in the following as "case $\alpha=4.0$" and to the second as case "case $\alpha=3.1$". Comparison between Pareto and exponential distribution makes sense of course only for $\alpha>3.0$ where Pareto distribution has finite standard deviation. First we investigate the difference in $\delta$-distribution function induced by different types of single claim size distributions. In Figure 7 those differences are depicted from 5% to 95% percentiles: "Diff" (black solid line; mean of 10 MC simulations) as well as standard deviation of 10 MC simulations in Pareto case ("Pstd"; stars) and exponential case ("Estd"; crosses). To get a better impression of the impact of claim number process on results we varied run-off period: I=5 (figure 7 a and b); I=20 (figure 7 c and d). For all examples we have chosen an exponential run-off pattern and a Poisson claim number process with $\lambda=100$. We tested for different run-off patterns and $\lambda$ and found similar results to those which are presented here. We define $Diff := F_{Expon.}^{-1} - F_{Pareto}^{-1}$; where F stands for $\delta$-distribution function. As Diff > 0 for small percentiles, we see from figure 7 that $\delta$-distribution function shows generally a higher risk for a too low reserving level in Pareto case than for exponential case, which is in accordance with values of single claim size standard deviation as given above. For the case $\alpha=4.0$ we observe no significant deviation between the two types of single claim size distribution because "Diff" lies below (at least one) standard deviation of MC simulation (see figure 7 a and c). On the contrary for $\alpha=3.1$ we find significant deviation at the left (high risk for underreserving) side of distribution functions (figure 7 b and d). This means single claim size distributions with large standard deviation generate a higher risk for insufficient CL reserves. This effect is enhanced for small run-off period as shown by a comparison between figures 7 b and d. Except figure 7 c we find higher skew for Pareto case than for exponential case, because "Diff" is higher on the left side than on the right. This means a higher risk for underreserving in Pareto case. Both types of single claim size distributions become comparable for high $\alpha$ (small standard deviation of claim sizes) and high claim numbers (long run-off periods) as shown by figure 7 c. For E(δ) we observe the following: In exponential case E(δ) is below zero as for all other simulations, too. E(δ) for Pareto case is always more negative than for exponential case. For $\Delta E(\delta) := |E_{Pareto}(\delta) - E_{Expon}(\delta)|$ we find values from 1% to 3% of $\sqrt{m.s.e.}$ for α=4.0 and from 5% to 10% of $\sqrt{m.s.e.}$ for α=3.1 depending on claim number process' parameters. This means δ-expectation's deviation from zero and therefore relative bias depend on the type of single claim size distribution, which is a reasonable behaviour. Next we investigate convergence properties of E(δ) by the use of exponential single claim size process for growing expected number in run-off table (Δ). We look at the case μ = 0.002 as for this case we expect better convergence properties than for the other due to the results presented before. In table 2c values for E(δ) are presented. A comparison with figure 3 and 6 shows that convergence properties for the exponential case are slightly better than for the Pareto case, but worse than for pure claim number case. We find no convergence E(δ) -> 0 until Δ < 5000 for the exponential case.

7. Conclusion

For all our simulations we find equal probability of over- and underreserving by the use of the CL method. But we also observe a negative relative bias in all cases. This means that the expectation of the difference between CL reserve and real reserve in terms of $\sqrt{m.s.e.}$ is negative. This result is of course restricted to the used claim model and has no general validity. On the other hand one has to take into account that the model in use is generally accepted for claim modelling and we investigate different types of distributions in large, relevant areas of parameter space.

In the case of the pure claim number model we find convergence to $E[\delta] \to 0$ for growing claim numbers in run-off triangle. Convergence is reached for approximately 5000 claims. For combination of claim number and claim size process according to the collective model we find decreasing relative bias with growing number of claims in run-off triangle but due to a saturation effect zero relative bias is not reached in our MC simulations. We observe several dependencies on distribution types and parameters which are quite reasonable.

We managed to establish a relation between the CL standard error and the risk of taking a too small reserve level. Different examples for results of the method developed in this paper are given. It is appropriate to measure safety loading in units of CL standard error as long as single claim size distribution has a finite second moment. CL reserves show a great risk of underreserving for single claim size distributions with infinite second moment. In this case standard error is not a proper measure for safety loading.

We conclude: Run-off data have to be checked very carefully, whether they fulfil the assumptions of the certain models used to justify a reserving method. If conditions for the CL method do not hold, it is very important to be aware of the loss data's generating process in order to calculate a sufficient reserve.

Acknowledgement

The author gratefully acknowledges discussions on the subject of this paper with Prof. K.D. Schmidt as well as the interesting and helpful comments of two referees.

## References

Hachemeister, C. A. and Stanard, J. N. (1975), IBNR Claims Count Estimation with Statistic Lag Functions, Paper Presented to the *XII<sup>th</sup> ASTIN Colloquium,* Portimao, Portugal

Institute of Actuaries (1997), *Claims Reserving Manual*, Second Edition, London

Mack, Th. (1991) A simple parametric model for rating automobile insurance or estimating IBNR claims reserves, *ASTIN Bulletin* **21**, 93 – 109

Mack, Th. (1993) Distribution – free Calculation of the Standard Error of Chain Ladder Reserves Estimates, *ASTIN Bulletin* **23**, 213 – 225

Mack, Th. (1997) *Schadenversicherungsmathematik*, DGVM Schriftenreihe angewandte Versicherungsmathematik Heft 28, Karlsruhe

Schmidt, K. D. and Wünsche, A. (1998) Chain Ladder, Marginal Sum and Maximum Likelihood Estimation, *DGVM – Blätter* **XXIII**, 267 – 277

Taylor, G. C. (1986) *Claims Reserving in Non-Life Insurance,* North-Holland, Amsterdam

Taylor, G. C. (2000) *Loss Reserving,* Kluwer Academic Publishers, Boston

Magda Schiegl

Cologne University of Applied Sciences

Claudiusstr. 1

D – 50678 Cologne

magda.schiegl@fh-koeln.de